\documentclass[prl,aps,twocolumn,floats,nofootinbib,showpacs]{revtex4}

\usepackage{amsmath,amssymb,dsfont,graphicx,psfrag}

\newcommand{\w}{\omega}

\begin{document}
\title{Large violation of Wiedemann Franz law in Luttinger liquids}

\author{Arti Garg$^{1}$, David Rasch$^{2}$, Efrat Shimshoni$^{3}$ and Achim Rosch$^{2,4}$}
\affiliation{$^{1}$Department of Physics, Technion, Haifa 32000, Israel\\
$^{2}$
Institute for Theoretical Physics, University of Cologne, 50937 Cologne, Germany\\
$^{3}$Department of Physics, Bar-Ilan University, Ramat-Gan 52900, Israel \\
$^{4}$ Kavli Institute for Theoretical Physics, University of California,
Santa Barbara, CA, USA}

\date{\today}
\begin{abstract}
We show that in weakly disordered Luttinger liquids close to a commensurate filling
the ratio of thermal conductivity $\kappa$ and electrical conductivity $\sigma$ can
deviate strongly from the Wiedemann Franz (WF) law valid for Fermi liquids scattering
from impurities. In the regime where the Umklapp scattering rate $\Gamma_U$ is much
larger than the impurity scattering rate $\Gamma_{\rm imp}$, the Lorenz number
$L=\kappa/(\sigma T)$ rapidly changes from very large values,
$L \sim \Gamma_U/\Gamma_{\rm imp} \gg 1$ at the commensurate point to very small values,
$L \sim \Gamma_{\rm imp}/\Gamma_{U} \ll 1$ for a slightly doped system. This surprising
behavior is a consequence of approximate symmetries existing even in the presence of
strong Umklapp scattering.
\end{abstract}

\maketitle

In a Fermi liquid, a quasi particle carries charge $e$ and has an energy of the order of $k_B T$.
These basic properties are reflected in the Wiedemann--Franz (WF) law \cite{wiedemann,sommerfeld}: the ratio of
the thermal conductivity divided by the
temperature $T$
and the electrical
conductivity, the so-called Lorenz number,
\begin{equation}
L = \frac{\kappa}{\sigma T}=\frac{\pi^2 k_B^2}{3 e^2} = L_0
\label{WF}
\end{equation}
takes a universal value $L_0$.
%which depends only on the Boltzmann constant $k_B$ and the electron charge $e$.
The WF
law, $L=L_0$, is valid and routinely observed
in the low-$T$ regime of Fermi liquids where impurity scattering dominates.
% and inelastic scattering processes can be neglected.

Deviations from the WF law, $L/L_0 \neq 1$,  in the low-$T$ regime, which
have e.g. been reported for high-temperature superconductors
\cite{hill} or
close  to quantum-critical points
\cite{paglione},  are regarded as evidence that the low-energy excitations cannot be viewed as electronic quasi particles.
But even if a description of thermal and electric transport in terms of Fermi liquid quasiparticles is possible,
the WF law will not be valid if inelastic scattering processes dominate which in general
relax heat- and charge currents differently. Typically, these corrections to $L/L_0$ are of the order of 1 and not
very large \cite{orignac,moreWF}.

Large violations of the WF law usually reflect a dramatic change of the excitation spectrum associated with the opening of a gap.
For example, in a Mott insulator $\sigma$ is exponentially small while heat can still efficiently be
transported by spin fluctuations. The opposite case occurs in a superconductor where $\sigma=\infty$  while $\kappa$
remains
finite at finite $T$ due to thermally excited
quasi particles.

\begin{figure}
\includegraphics[width=0.89 \linewidth]{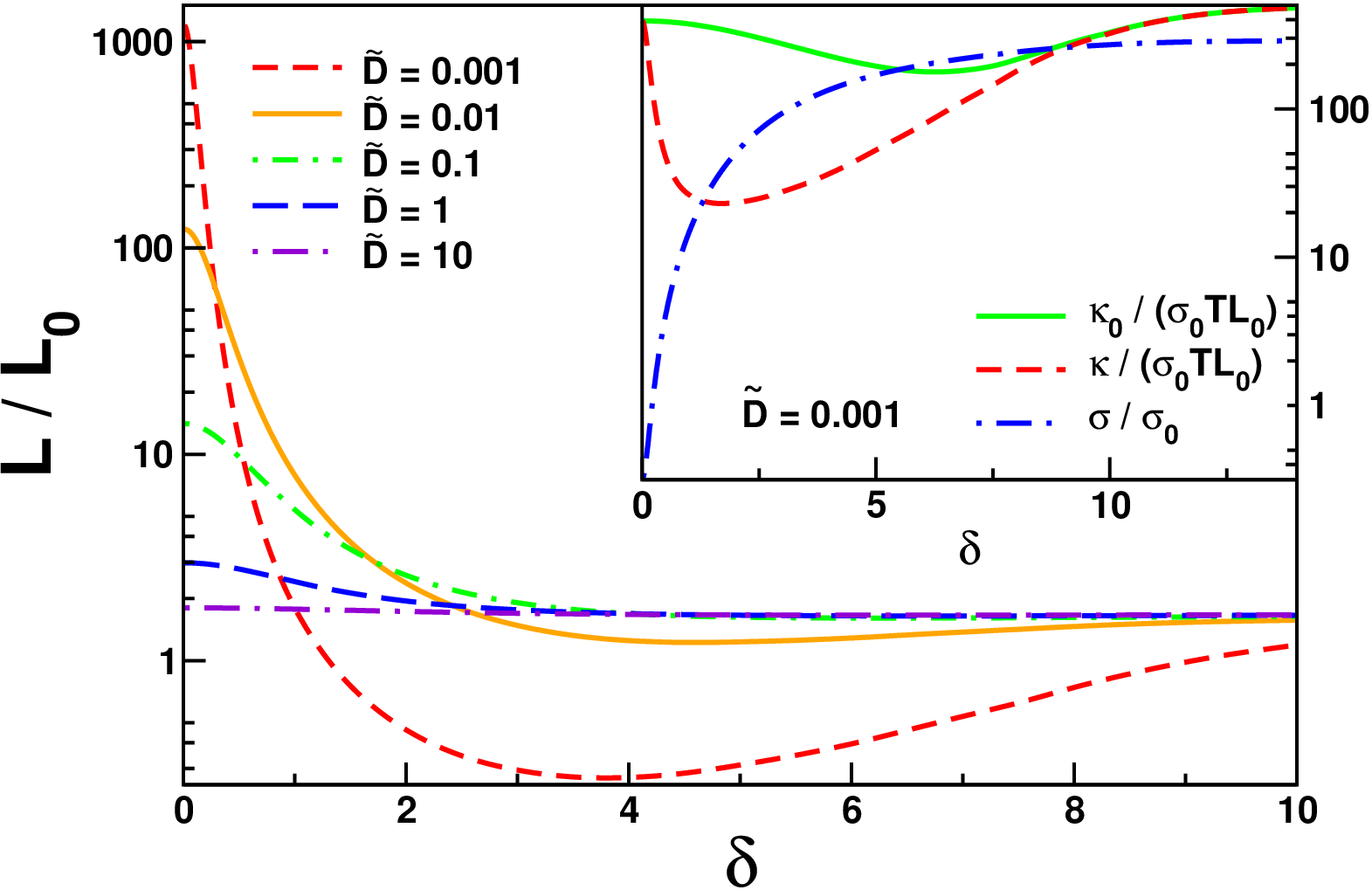}
\caption{
 Lorenz number, $L/L_0$,  (\ref{WF}) as a function of doping $\delta \nu$ away from 1/3
 filling ($\delta= 3 v_c G \delta \nu/(\pi T)$), using the variables of Eq. (\ref{dimless})
 (for $K_c=0.6, K_s=0.8, v_s/v_c=0.5$).
 If disorder dominates, $\tilde D\gtrsim 1$, $L/L_0$ is of order one and doping independent.
 For a clean system with $\tilde{D}\ll 1$, the WF law is strongly violated.
 A pronounced peak of height $1/\tilde{D}$ and width $\sqrt{\tilde{D}}$ at the commensurate
 filling is followed by a pronounced minimum. % at $\delta \sim 1$ of size $\tilde{D}$.
 Inset: $\delta$ dependence of $\kappa_0/(TL_0\sigma_0(T))$, $\kappa/(T L_0\sigma_0(T))$ and $\sigma/\sigma_0(T)$  for $\tilde{D}=0.001$, $\sigma_0(T)=(v_c^2a^{2n_c-3}/g^2)(v_c/aT)^\beta$ with $\beta=K_s n_s^2+K_c n_c^2-3$. \label{fig1} }
\end{figure}

In this paper, we show that small changes in the doping can trigger enormous changes of the Lorenz number $L$ in Luttinger liquids
in situations where the
Umklapp scattering rate $\Gamma_U$ is larger than the impurity scattering rate, $\Gamma_{\rm imp}\ll \Gamma_U$, see Fig.~\ref{fig1}. This happens even in regimes where
Umklapp scattering does {\em not} open a charge gap. This peculiar behavior can be traced back to the presence of approximate symmetries of the clean system which affect charge- and heat current in a completely different way.
This has to be contrasted with a situation where impurity scattering provides the dominant relaxation mechanism for both heat- and charge currents. For this case Li and Orignac \cite{orignac} have
shown that only violations of order $1$ of the WF law exist.

When investigating the thermal or electrical conductivity of low-dimensional systems,
 it is important to account for the role of symmetries and conservation laws even if these
 are only approximate. For example in integrable one-dimensional models, conductivities are usually infinite
 at finite $T$ \cite{reviewZotos} as the conservation laws protects the currents from decaying. Small perturbations
  render the conductivity finite, but still large \cite{almostIntegrable}.
 Below we demonstrate the implications on the thermoelectric effects.

We consider a weakly disordered one-dimensional (1D) metal described by a single band with the filling
$\nu=\nu_0+\delta \nu$, and the electron density $2 \nu$, where
$\nu_0=m_c/n_c$ with integers $m_c$, $n_c$ is a commensurate filling.
The low-energy Hamiltonian is given \cite{book} by
\begin{eqnarray}
H&=&H_{LL}+H_U+H_{\rm imp}\\
H_{LL}&=&  \int \frac{d x}{2 \pi} \sum_{i=c,s} v_i \left( K_i (\partial_x \theta_i)^2 +\frac{1}{K_i}  (\partial_x \phi_i)^2\right) \nonumber\\
H_U &=&\frac{g}{\left(2 \pi a\right)^{n_c}} \int dx \,  e^{i \sqrt{2}\left(n_c \phi_c\left(x\right)+n_s \phi_s\left(x\right)\right)} e^{-i \Delta k x} + h.c.
\nonumber
\\
H_{\rm imp}&=&
\frac{1}{\pi a}\int dx\,\eta(x) \left(e^{i\sqrt{2} \phi_c\left(x\right)}\cos\left(\sqrt{2}\phi_s\left(x\right)\right)+ h.c.\right) \nonumber
\end{eqnarray}
where $H_{\rm LL}$ is the usual Luttinger liquid Hamiltonian expressed in terms of spin (s) and charge (c) densities
 $\partial_x \phi_{c,s}$ and their conjugate variable $\partial_x \theta_{c,s}$ with
 $[\phi_{c,s}(x),\partial_{x'} \theta_{c,s}(x')]=i\pi\delta(x-x')$.  $H_U$ is the dominant Umklapp scattering process where
 $\Delta k = 2 n_c k_F - m_c G=n_c G \delta \nu$ (with $G=\frac{2\pi}{a}$) is proportional
to the deviation from commensurate filling and $n_s=0,1$ for even and odd $n_c$, respectively. The term $H_{\rm imp}$ with a
Gaussian correlated impurity
potential, $\langle \eta(x) \eta(x') \rangle=D \delta(x-x')$, describes
a weak backscattering due to disorder.

Even in the presence of Umklapp scattering, an approximate symmetry closely related to momentum conservation
exists \cite{pseudo}.  The so-called  pseudo momentum
\begin{eqnarray}
\tilde{P}&=&P_t-\frac{m_c G}{2 n_c} (N_R-N_L) =P+\frac{\Delta k}{2 n_c} (N_R-N_L)\  \ \label{pseudo}
\end{eqnarray}
(where $N_{R(L)}$ is the number of right(left) movers) commutes with $H_{LL}+H_U$ (even if effects like band curvature or a weak three-dimensional coupling are added \cite{pseudo,FL}). Here $P_t$ is the crystal momentum and $P=P_t-k_F(N_R-N_L)$  measures the momentum
relative to the two Fermi points.

%Due to the  pseudo momentum conservation, even a strong Umklapp scattering  can not equilibrate the system completely. Therefore the question whether $\kappa$ and $\sigma$ are proportional to $1/\Gamma_U$ or $1/\Gamma_{\rm imp}$ depends on the interplay
%of  charge and heat current, $J_c$ and $J_h$, and  $\tilde{P}$.

Because of the pseudo momentum conservation, even a strong Umklapp
scattering  may not be sufficient to relax the heat and charge
currents. To capture this, one needs a transport theory which
properly accounts for the role of conservation laws and the
associated vertex corrections. For the non-linear interaction
describing Umklapp scattering in Luttinger liquids the memory
matrix approach to transport \cite{forster} is to our knowledge
the only available method, especially as there are presently no
numerical methods to calculate conductivities at finite but low
$T$. As discussed in Ref.~\cite{bounds}, this method allows to
calculate  lower bounds to $\sigma$ and $\kappa$ in the
perturbative regime, and gives precise results as long as the
relevant slow modes are included in the calculation. It was shown to
capture prominent features of observable transport phenomena, e.g.
magnetothermal transport in spin-chains \cite{thermomagnetic}.

The first step to set up the memory matrix formalism, is to list a number of relevant operators $J_i$
which in our case includes the electrical current $J_1=J_c=v_c K_c (N_R-N_L)$, the heat current
$J_2=J_h=-\sum_{i=c,s} \int v_i^2  \partial_x \phi_i \partial_x \theta_i$ and the momentum operator $J_3=P=-
\sum_{i=c,s} \int   \partial_x \phi_i \partial_x \theta_i$. To leading order in $H_U$, $H_{imp}$, the
 matrix of conductivities is then obtained from
\begin{eqnarray}
\hat{\sigma}&=&\hat{\chi} \hat{M}^{-1} \hat{\chi}, \quad
M_{ij}=\lim_{\w \to 0} \frac{{\rm Im} \langle \partial_t J_i ; \partial_t J_j \rangle_\w}{\w} \label{mem}
\end{eqnarray}
with the $3 \times 3$ memory matrix $\hat{M}=\hat{M}_U+\hat{M}_{\rm imp}$.
%given by
%\begin{eqnarray}\label{mem}
%M_{ij}=\lim_{\w \to 0} \frac{{\rm Im} \langle \partial_t J_i ; \partial_t J_j \rangle_\w}{\w}\; .
%\end{eqnarray}
As the time derivatives $\partial_t J_i=i [H,J_i]$ are already linear in the weak perturbations $g_U$ and $\eta$,
the correlators are evaluated with respect to $H_{\rm LL}$. $\hat \chi$ is the matrix of static susceptibilities $\chi_{ij}=
\langle J_i;J_j \rangle_{\w=0}$ with
\begin{eqnarray}
\hat{\chi} &\approx& \frac{\pi T^2}{3} \left( \begin{array}{ccc}
\frac{6 v_c K_c}{\pi^2 T^2} & 0 & 0 \\
0 & v_c+v_s  &   \frac{1}{v_c}+\frac{1}{v_s}\\[1mm]
0 &   \frac{1}{v_c}+\frac{1}{v_s} &  \frac{1}{v_c^3}+\frac{1}{v_s^3}
\end{array}
\right)\; .
\end{eqnarray}
The Umklapp contribution to Eq.~(\ref{mem}) is given by
%\begin{widetext}
%\begin{eqnarray}
%\hat{M}_U &\approx& \frac{g^2 (\pi T a)^{K_cn_c^2+K_sn_s^2-1}}
%{(2 \pi a)^{2 n_c-1} v_c^{K_c n_c^2-1}v_s^{K_sn_s^2} } \\
%&& \times
%\left( \begin{array}{ccc}
%\frac{4 v_c^2 n_c^2K_c^2}{\pi T^2}F_{00} &
%\frac{2\pi v_c n_c K_c }{ \Delta k} F_{3} &
%-\frac{2 v_cn_cK_c  \Delta k}{ \pi T^2}F_{00} \\
% \frac{2\pi v_c n_c K_c }{ \Delta k} F_{3}&
%-\pi  v_c^2 F_{4} &
%\pi   F_{3} \\
%-\frac{2 v_cn_cK_c  \Delta k}{\pi T^2}F_{00}  &
% \pi F_{3} &
%\frac{ \Delta k^2}{\pi T^2}F_{00}
%\\ \end{array}\right)\nonumber
%\end{eqnarray}
%\end{widetext}
\begin{eqnarray}
\frac{\hat{M}_U}{c_U \Gamma_U} &\approx& \left(
\begin{array}{ccc} \frac{2 v_c^2 n_c^2K_c^2F_{00}}{\pi T^2} &
\frac{ v_c n_c K_c F_{3}}{ \Delta k} &
\frac{- v_cn_cK_c  \Delta k F_{00}}{ \pi T^2} \\
 \frac{ v_c n_c K_c F_{3}}{ \Delta k} &
-v_c^2 F_{4}/2 &
  F_{3}/2 \\
\frac{- v_cn_cK_c  \Delta k F_{00}}{\pi T^2}  &
  F_{3}/2 &
\frac{ \Delta k^2 F_{00}}{2\pi T^2}
\\ \end{array}\right)~~~
\end{eqnarray}
where $c_U = \frac{(\pi)^{K_c n_c^2+ K_s n_s^2-1}}{(2 \pi)^{2n_c-1}}\left(\frac{v_c}{v_s}\right)^{K_sn_s^2}$
and $\Gamma_U = \frac{g^2}{a^{2n_c-1}}\left(\frac{aT}{v_c}\right)^{K_c n_c^2+K_s n_s^2-1}$. $F_{mn}$ are the dimensionless functions
\begin{eqnarray}
F_{mn}&=&2\int dx dt ~\,t\,e^{i\delta x} (\partial_x^m f_c(x,t)) (\partial_x^n f_s(x,t))\\
f_c(x,t)&=&\left(\sinh(x+it)\sinh(x-it)\right)^{-\frac{K_cn_c^2}{2}} \nonumber \\
f_s(x,t)&=&\left(\sinh\left(x v_c/v_s +it\right)\sinh\left(x v_c/v_s -it\right)\right)^{-\frac{K_sn_s^2}{2}}\nonumber \\
F_{3}&=& \pi[F_{20}+(v_s/v_c)^2 F_{02}  +(1+(v_s/v_c)^2)F_{11}] \nonumber\\
F_{4}&=&  \pi[F_{20}+(v_s/v_c)^4 F_{02}+ 2 (v_s/v_c)^2 F_{11}] \;
,\nonumber
\end{eqnarray}
%with $\langle\phi_c(x)\phi_c(0)\rangle=K_c \ln .......................$ and $\langle\phi_s(x)\phi_c(0)\rangle=K_s \ln %................$
which depend on  doping and $T$ via $\delta=v_c \Delta k / (\pi T)$. Note that
$\hat{M}_U$ has a vanishing eigenvalue %with eigenvector $(-\Delta k/(2 v_c K_c),0,1)$
reflecting that $[H_U,\tilde{P}]=0$.
The disorder contribution is given by
%\begin{widetext}
\begin{eqnarray}
\frac{\hat{M}_{\rm imp}}{c_{\rm imp} \Gamma_{\rm imp}} &\approx&
\left(
\begin{array}{ccc} \left(\frac{4 K_c v_c}{2\pi  T}\right)^2& 0&
0\\
0& %\frac{(K_cv_c^2+K_sv_s^2)K_t}{v_c v_s (1+K_t)}
v_c v_s \tilde{K}
&\frac{K_t^2}{1+K_t}\\
0&
\frac{K_t^2}{1+K_t}&
 \frac{(\frac{K_c}{v_c^2}+\frac{K_s}{v_s^2})K_t}{ 1+K_t}
\end{array}\right)
\end{eqnarray}

%D \frac{(2\pi T a)^{K_t}}{4 \pi a^2 v_c^{K_c} v_s^{K_s}} \frac{\Gamma^2(K_t/2)}{\Gamma(K_t)}
% \\
%&& \times \left( \begin{array}{ccc}
%\left(\frac{4 K_c v_c}{2\pi  T}\right)^2&
%0&
%0\\
%0& %\frac{(K_cv_c^2+K_sv_s^2)K_t}{v_c v_s (1+K_t)}
%v_c v_s \tilde{K}
%&\frac{K_t^2}{1+K_t}\\
%0&
%\frac{K_t^2}{1+K_t}&
% \frac{(K_c/v_c^2+K_s/v_s^2)K_t}{ 1+K_t}
%\end{array}\right)\nonumber
%\end{eqnarray}
%\end{widetext}
where $c_{\rm imp} = \frac{(2\pi)^{K_t-1}}{2}\left(\frac{v_c}{v_s}\right)^{K_s} \frac{\Gamma^2(K_t/2)}{\Gamma(K_t)}$,
$\Gamma_{\rm imp} = \frac{D}{a^2}\left(\frac {aT}{v_c}\right)^{K_t}$, $K_t=K_c+K_s$ and $\tilde{K}=\frac{(K_cv_c^2+K_sv_s^2)K_t}{v_c v_s (1+K_t)}$. Finally,
 $\sigma$,  $\kappa$ and $L$  of Eq.~(\ref{WF}) are obtained from
 \begin{eqnarray}
\sigma&=&\hat{\sigma}_{11}, \quad \kappa=\kappa_0-T S^2 \sigma=
\frac{1}{T} \left(\hat{\sigma}_{22}-\frac{\hat{\sigma}_{21}^2}{\hat{\sigma}_{11}} \right). \label{te}
\end{eqnarray}
It should be noted that $\kappa$ is measured experimentally
in a setup where the charge current vanishes, resulting
in the thermoelectric counter terms of Eq.~(\ref{te}).
% which are
%subtracted from the $J_h$-$J_h$ correlation function $\kappa_0$.
$S=\hat\sigma_{21}/(T \hat\sigma_{11})$ is the thermopower.

For given Luttinger liquid parameters $v_{c,s}, K_{c,s}$, the Lorenz number depends only on two dimensionless quantities,
describing the ratio of renormalized disorder strength and Umklapp scattering and the doping:
\begin{eqnarray}\label{dimless}
\tilde{D}=\frac{\Gamma_{\rm imp}}{\Gamma_U}=\frac{D a^{2 n_c-3}}{g^2 (a T/v_c)^{\gamma}}, \quad \delta=\frac{v_c \Delta k}{\pi T}
\end{eqnarray}
with $\gamma=
(n_c^2-1) K_c+(n_s^2-1) K_s-1$.
Fig.~\ref{fig1} shows the striking doping dependence of $\sigma, \kappa$ and
the Lorenz number $L/L_0$ for the filling $1/3$ ($n_c=3$, $n_s=1$).
For large effective disorder, $\tilde{D}\gtrsim 1$, $L/L_0$ is of order 1 and there is essentially no doping dependence.
For $\tilde D\ll 1$ one obtains instead a huge and sharp peak of height $1/\tilde D$ and width $\sqrt{\tilde D}$ followed by a wider dip
located at $\delta \sim 1$, where the minimum scales as $\tilde D$.

This behavior can be understood by investigating the relation of the currents $J_h$ and $J_c$ to the approximately
conserved $\tilde P$, Eq. ~(\ref{pseudo}). From the continuity equation, one can show \cite{FL}
that the cross susceptibility of $J_c$ and $\tilde{P}$ is (up to exponentially small corrections) given
by the doping $\delta \nu$ away from the commensurable point
\begin{eqnarray}
\chi_{J_c \tilde{P}}&=& 2\delta \nu \approx \frac{\Delta k \hat\chi_{11}}{2 n_c K_c v_c} +\hat\chi_{31}  \label{overlap}
\end{eqnarray}
while $\chi_{J_h \tilde{P}}\sim T^2 >0$.
%For the heat current $J_h$ we obtain
%\begin{eqnarray}
%\chi_{J_h \tilde{P}} \approx \frac{\Delta k \hat\chi_{12}}{2 n_c K_c v_c} +\hat\chi_{32}=\frac{\pi T^2}{3} \left(\frac{1}{v_c}+\frac{1}{v_s} \right) >0\; .
%\end{eqnarray}
$\chi_{J_i \tilde{P}}$ measures the ''overlap`` of the current and the conserved operator.
A vanishing $\chi$ implies that the operators are orthogonal
to each other, i.e. the current is {\em not} protected by the conservation law
and can decay rapidly by Umklapp processes.
Therefore, at the commensurate point where $\chi_{J_c \tilde{P}}=0$, $J_c$ can decay
by Umklapp processes, while $J_h$ is protected. Indeed,
as shown in the inset of Fig.~\ref{fig1}, at $\delta=0$ one obtains $\sigma\sim 1/\Gamma_U$
small, but $\kappa\sim 1/\Gamma_{\rm imp}$, resulting
in $L/L_0 \sim 1/\tilde D$ in the clean limit, $\tilde D \ll 1$.

For finite doping, $\chi_{J_c \tilde{P}}=\delta \nu >0$ and therefore
$\sigma\sim (\delta \nu)^2/\Gamma_{\rm imp}$ grows rapidly until it becomes of the same order as the heat conductivity
in the {\em absence} of electrothermal correction, $\kappa_0/T$. In this regime, the leading contribution to
 $\kappa/T$, however, of order $1/\Gamma_{\rm imp}$ is exactly canceled by the thermoelectric counter terms
 in Eq. (\ref{te}). The physical origin of this cancelation is that $\kappa$ is measured under the boundary condition
 $J_c=0$. As the component of $J_c$ perpendicular to $\tilde P$ decays rapidly by Umklapp,
 $J_c$ and $\tilde P$ become almost parallel for small $\tilde D$ implying that effectively the
 heat conductivity measurement
 is performed under the boundary condition of vanishing $\tilde P$. Therefore $\kappa$ becomes of order $1/\Gamma_U$,
 and $L/L_0\sim\tilde D$.
 % (times a rather large prefactor).
 For neutral liquids a related effect is well known: while mass
 currents do not decay due to momentum conservation, the heat conductivity measured under the boundary
 condition of vanishing
 mass currents remains finite (this situation is more transparent as momentum and mass
  current are proportional to each other
 while this is not the case for $J_c$ and $\tilde P$). Finally, for $\delta \gg 1$ the Umklapp scattering
 is exponentially suppressed, both $\sigma$ and $\kappa/T$ are of order $1/\Gamma_{\rm imp}$, and $L/L_0\sim 1$ \cite{orignac}.

\begin{figure}
\includegraphics[width=0.88 \linewidth,clip]{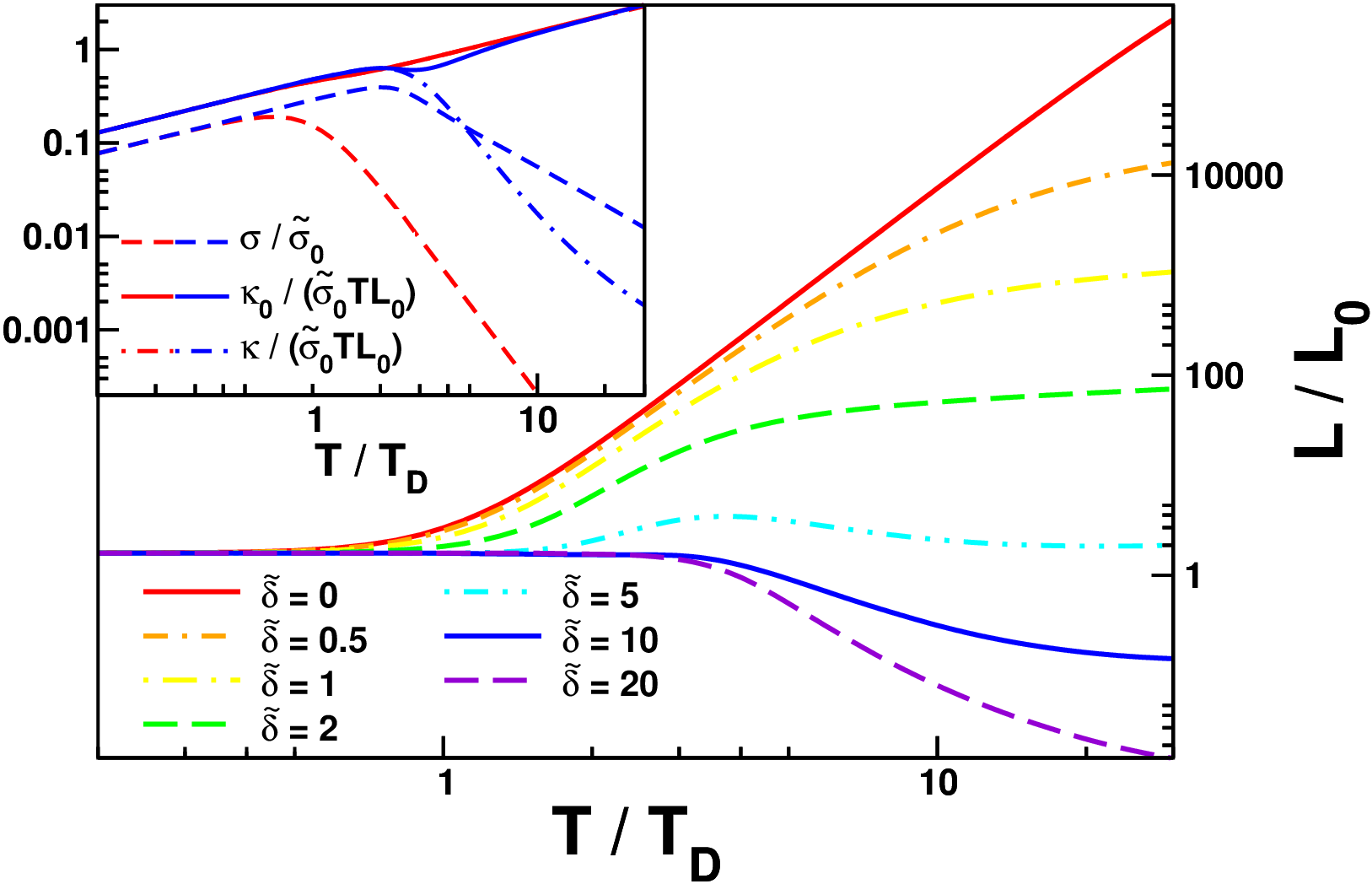}
\caption{
$T$ dependence of the Lorenz number for various dopings close to $1/3$ filling using (\ref{dimless2})
(parameters as in Fig.~\ref{fig1}).  At low $T$ disorder always dominates resulting in a $T$-independent
$L/L_0$ of order 1. At the commensurate point $L/L_0 \sim 1/\tilde{D}$.
Inset: $\kappa_0(T)/(TL_0\tilde{\sigma}_0(D))$, $\kappa(T)/(TL_0\tilde{\sigma}_0(D))$ and $\sigma(T)/\tilde{\sigma}_0(D)$ for
$\tilde{\delta}=0$ (red) and $\tilde{\delta}=10$ (blue).
Here $\tilde\sigma_0(D)= [Da^{2n_c-3}/g^2]^{\alpha} v_c^2/D$ with $\alpha= (2-K_c-K_s)/\gamma$.
}
\label{fig2}
\end{figure}

In Fig.~\ref{fig2} the $T$ dependence of the WF ratio, $\sigma$ and $\kappa$ are shown using the appropriate dimensionless variables
\begin{eqnarray}\label{dimless2}
\tilde\delta&=&\frac{\delta}{{\tilde D}^{1/\gamma}}, \quad
\tilde T=\frac{T}{T_D}, \quad T_D\equiv\frac{v_c}{a} \left(\frac {D a^{2 n_c-3}}{g^2} \right)^{1/
\gamma}.
\end{eqnarray}
Upon lowering $T$, the disorder close to $1/3$ filling becomes more and more important, $\tilde{D}$ grows  and $L/L_0$ becomes
 of order $1$ for low $T$. As explained above, for vanishing doping $\tilde \delta=0$, $\sigma$ is much smaller than $\kappa/T$
 as long as Umklapp scattering dominates. For finite doping,
 Umklapp scattering is exponentially suppressed at low $T$ (see inset of Fig. \ref{fig2}). However when it sets in ($\tilde T > 1$), it leads
 to a larger suppression of $\kappa/T$ compared to $\sigma$ due to the partial cancellations
 from thermoelectric corrections.

\begin{figure}
\includegraphics[width=0.88 \linewidth,clip,angle=0]{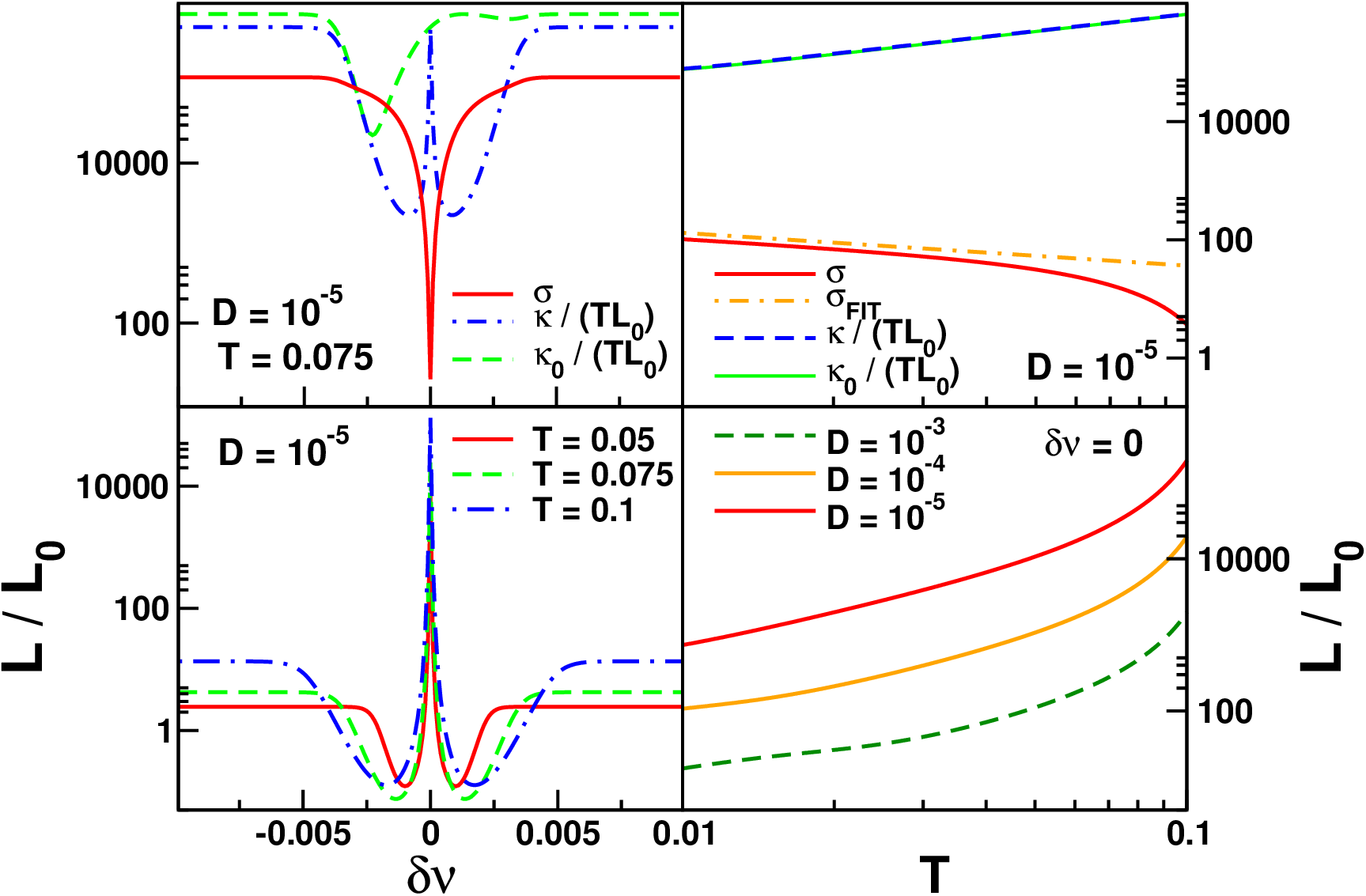}
\caption{Lorenz number (lower curves), $\kappa$ and $\sigma$ (upper curves) for a system close to
$1/4$--filling, ($\sigma_{\rm FIT} = 10T^{-0.56}$ is the fit to $\sigma$) where $K_c=0.22$ (chosen to be compatible with Ref.~\cite{bechgaard}),
$K_s=0.8, v_s/v_c=1/2$, $g=0.1v_c a^{n_c-3/2}$, $ n_c=4, n_s=0, m=1/v_ca$; $D$ is in units of $g^2/(a^{2n_c-3})$, and
$T$ (in units of $v_c/a$) is in the experimentally accessible regime.
%half filling where parameters ($K_c=0.6, K_s=0.8, v_s/v_c=1/2, n_c=2, n_s=0, m=1$)
%have been chosen
%so that $\tilde{D}=0.01$ is $T$ independent.
\label{fig3}}
\label{bandcurve_fig}
\end{figure}

While the theoretical analysis of the problem described above is
most transparent for the filling close to $1/3$, it is useful to
study a case with direct experimental realizations. One possible
candidate is the quarter-filled quasi-1D Bechgaard salt
(TMTSF)$_2$PF$_6$ \cite{bechgaard} where the anisotropy of the
kinetic energy ($t_a:t_b:t_c = 250:20:1$ meV) allows a Luttinger
liquid description for $T\gtrsim 100K$.
 Two extra complications arise at quarter filling: first, in the absence of disorder the effective
low-energy model, $H_{LL}+H_U$ becomes the integrable sine-Gordon
model, which formally has an infinite number of conservation laws
on top of the pseudo momentum. For an analysis of transport one
has to identify %and take into account
the leading corrections which break integrablity (see
Ref.~\cite{almostIntegrable}). Second,  for $H_{LL}+H_U$ there is
a strict separation of charge and spin degrees of freedom the
latter being not affected by Umklapp scattering. We therefore have
to take band-curvature \cite{haldane} into account, which couples
spin and charge and breaks integrability:
\begin{align}
H_{BC}=-\frac{1}{6\sqrt{2}m} \int \bigl(\partial_x\phi_c^3 + 6\partial_x\phi_s\partial_x\theta_s\partial_x\theta_c
\nonumber \\
+ 3\partial_x\phi_c
(\partial_x\phi_s^2 + \partial_x\theta_s^2 + \partial_x\theta_c^2)\bigr)- \delta \mu \int \partial_x\phi_c\; .
\end{align}
Here we have added an extra $T$-dependent chemical potential
$\delta\mu = \frac{T^2\pi^2}{12m}\left(\frac{1}{v_c^2}\left(K_c+K_c^{-1}\right)+\frac{1}{v_s^2}\left(K_s+K_s^{-1}\right)\right)$
%-\frac{1}{2\sqrt{2}m}\langle\partial_x\phi_c(x)^2+\partial_x\phi_s(x)^2+
%\partial_x\theta_c(x)^2+\partial_x\theta_s(x)^2\rangle$
to account for the $T$-independent particle density $2 \nu$ in a
3D crystal. To leading order in $1/m$, corrections to $\hat \chi$
arise only for $\chi_{12}=\chi_{21}\approx \frac{\pi T^2}{3 m}
 \left( 1/v_c+1/v_s \right)$ and $\chi_{13}=\chi_{31}\approx \frac{\pi T^2}{3 m}
 \left(1/v_c^3+1/v_s^3\right)$. As both $N_R-N_L$ and $P$ commute with $H_{BC}$,
 only $\hat{M}_{22}$ gets an extra contribution,
 $\hat{M}_{22}^{BC}=\frac{\pi^8T^5}{128m^2v_s^4v_c^4}K_c\left(K_s^{-2}+K_s^2-2\right)
 \int t {\rm Im}[(4\cosh^2(x+it)+2)\sinh(x+it)^{-4}\sinh(xv_c/v_s+it)^{-2}\sinh(xv_c/v_s-it)^{-2}]$.
 As $J_c \to J_c+P/m$, $\sigma$ is given by
 $\sigma=\hat \sigma_{11}+2 \hat\sigma_{13}/m+\hat{\sigma}_{33}/m^2$ (the corresponding correction to $J_h$
 is subleading and therefore omitted).

 An example for the expected doping and $T$ dependencies %for $\kappa/T$, $\sigma$ and $L/L_0$
 is shown in
 Fig.~\ref{fig3} for a filling close to $1/4$ using parameters consistent with existing resistivity data
 for (TMTSF)$_2$PF$_6$ \cite{bechgaard}. Both $\rho(T)$ and  $\sigma(\omega)$
 in this system can  be explained \cite{bechgaard} by Umklapp
 scattering in a $1/4$ filled Luttinger liquid with
$K_c\approx 0.22$ leading to $\rho \sim g^2 T^{16K_c-3}$ (i.e. $\sigma \sim T^{-0.56}$, see Fig.~\ref{fig3}) along the chain.
Other parameters like $K_s$, $v_s$, $m$,
and, most importantly, disorder strength $D$, are not known experimentally. The absence of any visible
disorder contribution to $\rho(T)$ in the Luttinger liquid regime, $T\gtrsim 100$K, allows us to estimate crudely
$D \ll 0.0005$ in units of $g^2/a^{2n_c-3}$. Our results shown in Fig.~\ref{fig3} strongly suggest that a large
 violation of the WF law (after subtraction of the phonon contribution not discussed here)
 should be observable in Bechgaard salts and similar materials.

Qualitatively, the doping dependence of $L/L_0$ for $1/4$ and $1/3$ filling are
 similar. The WF ratio $L/L_0$  shows
 a pronounced sharp peak of height $1/\tilde{D}$ followed by a dip for $v_c \Delta k \sim T$.
 %At low $T$, the dip
 %is, however, less pronounced as the scattering of the spin degrees of freedom due to band-curvature effects is
 %very weak, yielding a large $\kappa$.
% Remarkably, $\sigma$ and $\kappa$ are symmetric functions of the
% doping within our approximations,
% while $\kappa_0$ becomes asymmetric due to band-curvature effects.
 $T$-dependencies might differ in the two cases due to the different $T$ dependence of $\tilde{D}$:
whether $1/\tilde{D}$ grows or shrinks upon lowering $T$ depends
on $K_c$ and $K_s$.
 %In Fig.~\ref{fig3} they are chosen such
 %that $\tilde{D}$ is $T$ independent.
 However, the most prominent
 $T$-dependence arises from the fact that  Umklapp scattering is effectively switched off at lowest $T$ for $\delta \nu >0$,
 resulting in $L \sim L_0$.
 %(further $T$ dependencies arise from  band-curvature effects).

We expect that the strong violation of the WF law in regimes
where Umklapp scattering is large compared to disorder
will not only occur for the strictly 1D systems discussed here but even if a weak inter-chain
 tunneling (as in case of Bechgaard salts) is taken
 into account,
as a small modulation of the 1D bands  does not affect the structure of approximate
conservation laws, see \cite{FL}. %\cite{leads}.
%{\bf But in presence of interchain tunneling, there can be further
%complications because of incoherent scattering when the electron tunnels between different chains.
%These effects are known not to effect the dc resistivity much, but they can have some
% effect on heat conductivity,although we expect our main result still to hold.
%  These questions require further investigations.}
Besides the disparate behavior of $\kappa/T$ and $\sigma$ an interesting
finding of our study is the importance of thermoelectric corrections for the slightly doped system. In the regime where
$L/L_0$ gets very small due to a partial cancelation of $\kappa_0$ and $T S^2 \sigma$, the
dimensionless thermoelectric figure of merit, $Z T=T \sigma S^2/\kappa_0$,
which measures the efficiency of a thermoelectric element for power generation or refrigeration,
 becomes  $1$, a remarkably large value \cite{figureOfMerit}.

 This work was supported by the DFG under SFB 608, the NSF grant PHY05-51164
 and the German-Israeli Foundation (GIF).

\end{document}